\documentclass{PoS}

\title{Models of Optical Emission Lines To Investigate Narrow-Line Seyfert 1 Galaxies in Spectroscopic Databases}

\ShortTitle{Models of NLS1 Optical Emission Lines}

\author{\speaker{G. La Mura},$^a$ D. Bastieri,$^{ab}$ M. Berton,$^{cd}$ S. Chen,$^{af}$ S. Ciroi,$^a$ E. Congiu,$^{ae}$ V.~Cracco,$^a$ J.-H. Fan,$^f$ P. Rafanelli$^a$\\
  \llap{$^a$} Dipartimento di Fisica e Astronomia ``G. Galilei'', Universit\`a di Padova, Vicolo dell'Osservatorio 3, 35122, Padova, Italy\\
  \llap{$^b$} Istituto Nazionale di Fisica Nucleare (INFN), Sezione di Padova, 35131, Padova, Italy \\
  \llap{$^c$} Finnish Centre for Astronomy with ESO (FINCA), University of Turku, Quantum, Vesilinnantie 5, 20014 University of Turku, Finland \\
  \llap{$^d$} Aalto University Mets\"ahovi Radio Observatory, Mets\"ahovintie 114, FIN-02540 Kylm\"al\"a, Finland \\
  \llap{$^e$} INAF - Osservatorio Astronomico di Brera, Via E. Bianchi 46, 23807, Merate (LC), Italy \\
  \llap{$^f$} Center for Astrophysics, Guangzhou University, 510006, Guangzhou, China \\
  E-mail: \email{giovanni.lamura@unipd.it}
}

\abstract{Thanks to the execution of extensive spectroscopic surveys that have covered large fractions of the sky down to magnitudes as faint as $i \approx 19$, it has been possible to identify several narrow-line Seyfert 1 galaxies (NLS1s) and to investigate their properties over a large range of the electromagnetic spectrum. The interpretation of their nature, however, is still hampered by the statistical uncertainties related to the way in which NLS1 candidates are selected. In this contribution, we present a study to detect and to model emission lines in optical spectra extracted from the Sloan Digital Sky Survey (SDSS), adopting the most proper strategy to identify the source of line excitation and to produce a detailed model with measurements of several emission line parameters. We demonstrate the application of this technique to explore fundamental questions, such as the presence of gas and dust around the core of AGNs and the spectral energy distribution of their ionizing radiation. We compare the spectral properties of NLS1s with those of other type 1 active galaxies and we summarize the potential of this approach to identify NLS1s in present day and future spectroscopic surveys. We finally consider the implications of multi-frequency data analysis in the debate concerning the intrinsic nature of NLS1s.}

\FullConference{Revisiting narrow-line Seyfert 1 galaxies and their place in the Universe -  NLS1-2018\\
		9-13 April 2018 \\
		Padova Botanical Garden, Italy}

\begin{document}
\newcommand{\ion}[2]{#1~{\small #2}}

\section{Introduction}
The unusually small velocity fields, corresponding to less than $2000\, {\rm km\, s^{-1}}$ in full width at half maximum (FWHM), that characterize the optical spectra of narrow-line Seyfert 1 galaxies (NLS1s, \cite{Osterbrock85}), pose one of the most intriguing challenges to our understanding of active galactic nuclei (AGNs). A simple obscuring structure in front of the central regions of the source, which well explains most of the observational properties of broad-line type 1 AGNs and narrow-line type 2s \cite{Antonucci93}, does not apply to the case of NLS1s. Many of these AGNs show prominent \ion{Fe}{II} multiplets \cite{Laor97,Vestergaard01} and non-stellar continuum in the optical spectra, together with strong soft X-ray emission characterized by rapid variability (down to less than $1\,$hr, \cite{Boller96}). These clues suggest that the central source is not obscured. The low velocity fields of NLS1s are, therefore, interpreted in two ways: they may be either attributed to the gravitational attraction of a relatively low mass ($M \approx 10^7\, {\rm M}_\odot$) central super massive black hole (SMBH, \cite{Mathur00, Mathur01}), or they could be the result of extreme projection effects from a flattened rotating structure, seen nearly face-on \cite{Decarli08}.

Combining spectroscopic observations with the host galaxy properties, where it is well established, at least in the low redshift regime, that NLS1s reside in spiral hosts that should correspond to typically low mass black holes \cite{Orban11}, several lines of evidence support the low-mass interpretation over the geometrical one. Although it is demonstrated that geometrical effects play some role in the determination of the optical broad line profiles, NLS1s do not appear to be exceptional compared to other type 1 sources in this point of view \cite{LaMura09}. In addition, radio morphology studies suggest that we observe NLS1s at various orientations \cite{Berton18}. On the contrary, the presence of a low mass SMBH in their nuclei implies that NLS1s work at quite high accretion regimes, close to and sometimes beyond the Eddington limit, in order to account for their observed luminosity. This particular property would identify NLS1s with quickly evolving sources, which might be in the first stages of their SMBH growth history. As a consequence, they may provide fundamental hints about the influence of AGN activity in a cosmological context.

With this contribution, we present the results obtained from a set of models designed to reproduce the properties of optical emission lines in large spectroscopic databases, like the Sloan Digital Sky Survey (SDSS, \cite{York00}) or the 6dF Galaxy Redshift Survey (6dFGS, \cite{Jones04, Jones09}). We use a systematic sequence of models in order to extract emission-line parameters from spectra and to distinguish among the various classes of astrophysical sources with optical emission lines. We subsequently focus our analysis on the properties of type 1 AGNs and we use the resulting data sample to compare NLS1s with the broad-line emitting sources.

\section{Data selection and analysis}
In recent times, several authors have investigated the possibility of identifying NLS1s by modelling the profiles of the emission lines detected in the optical spectra published by extensive spectroscopic surveys \cite{Cracco16, Rakshit17, Chen18}. With the publication of the Fourteenth Data Release of SDSS (DR14, \cite{Abolfathi18}) and plans to extend the sky coverage of spectroscopic observations in the Southern Hemisphere, it has become clear that the possibility of modelling spectra and extracting emission-line parameters will be of fundamental importance in order to take advantage of the wealth of observational material that is being released to the scientific community. As a consequence, we started a project to collect data from multiple frequency archives and to combine them with optical spectra, with the aim of selecting large object samples and investigating their statistical properties.

\begin{figure}[t]
  \includegraphics[width=0.48\textwidth]{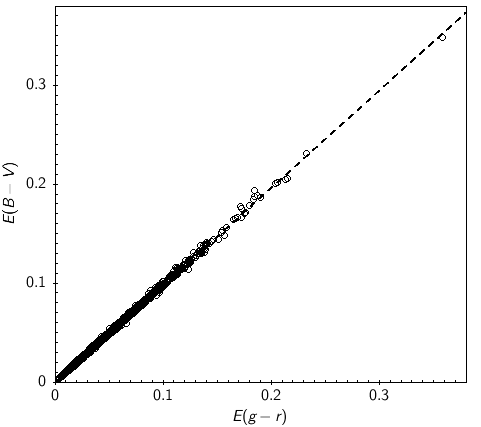}
  \includegraphics[width=0.46\textwidth]{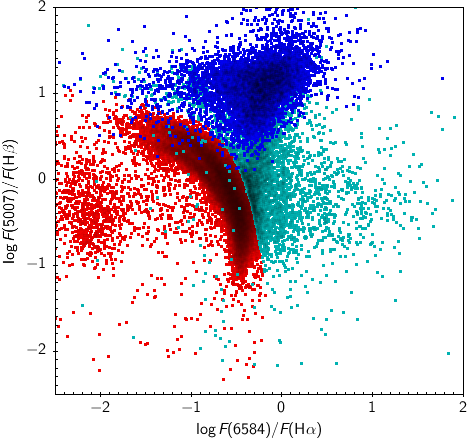}
  \caption{{\bf Left panel -} Relationship between the SDSS $E(g -r)$ color excess and the $E(B - V)$ color excess reported for the GALEX observation of 9158 QSOs. The dashed line represents the linear fit of Eq.~(\ref{eqExcess}), with a linear correlation coefficient $r > 0.99$. {\bf Right panel -} BPT diagram of 53951 sources, with succesfully fitted diagnostic lines. Here we represent as red dots objects that were fitted as star forming galaxies, with blue points type 1 AGNs and with cyan points type 2 AGNs. The narrow-line components of type 1 objects are consistent with the type 2 population.}
\end{figure}
By querying the summary of emission-line measurements of the SDSS database\footnote{For this purpose, we used the spectroscopic pipeline data products of SDSS, stored in files named \texttt{spZline-PPPP-MMMMM.fits} for every exposed plate.}, we selected all the spectra with measurements of the optical emission lines H$\beta$, [\ion{O}{III}]$\lambda\lambda 4959,5007$ and H$\alpha$ detected with a signal-to-noise ratio better than 10 and not classified as \texttt{STAR} by the spectroscopic pipeline. We adopted this strategy, instead of directly selecting spectroscopically identified QSOs, because some type 1 objects with prominent narrow line components may happen to be classified as \texttt{GALAXY} and therefore get lost in samples selected on the basis of the \texttt{QSO} classification. We obtained the spectra of 66423 line-emitting galaxies, for which we performed a series of diagnostic operations. First, we applied a Cardelli-Clayton-Mathis extinction correction (CCM, \cite{Cardelli89}), based on the $E(g-r)$ color excess returned by the SDSS photometric pipeline. We converted this into $E(B-V)$ through the empirical relation:
\begin{equation}
  E(B-V) = 0.984 \cdot E(g-r) \label{eqExcess}
\end{equation}
determined from 9158 QSOs, selected from a cross match of SDSS-DR14 with GALEX observations \cite{Bianchi11}, and illustrated in Fig.~1 (left). Then we transformed all the collected spectra to the rest frame, using the redshift measurement provided by SDSS. After these preliminary steps, we ran a first simple iterative fit to the H$\beta$\ region to infer whether the object has broad emission lines. If the total H$\beta$\ profile is estimated to be broader than [\ion{O}{III}]$\lambda 5007$, the object is treated as a type 1 AGN candidate. The Balmer lines are fitted with multiple Gaussians (1 narrow component, and 1 or 2 broad components, depending on the signal to noise ratio) and the forbidden lines of [\ion{O}{I}]$\lambda 6300$, [\ion{O}{I}]$\lambda 6364$, [\ion{N}{II}]$\lambda\lambda 6548,6584$ and [\ion{S}{II}]$\lambda\lambda 6717,6731$ are fitted with single Gaussians. The [\ion{O}{III}]$\lambda\lambda 4959,5007$ doublet represents a special case that we modelled with a narrow-core component and, in the high S/N spectra, an additional broad/shifted component, requiring, in this case, to fit both lines with the same profile. If no broad component is detected, the object is modelled as a type 2 AGN or a star-forming galaxy, and the emission lines are fitted with single Gaussians, but we still allow for the possible detection of a broad component in H$\alpha$\ (LINER, Seyfert 1.9). This procedure was designed to extract line flux and profile estimates, taking into account the complexity of the recombination lines in type 1 AGNs, which arises from the combination of several spectral contributions. It is able to extract emission line models for a total of 66097 objects. In order to test the consistency of the fitting procedure, we compared the distribution of the narrow line components in type 1 AGNs with the Baldwin-Phillips-Terlevich diagnostic diagram (BPT, \cite{Baldwin81, Veilleux87, Kewley06}) of star forming galaxies and type 2 AGNs. The right panel of Fig.~1 shows that the samples behave consistently, suggesting that the narrow line contributions are reasonably well identified.

\begin{figure}[t]
  \begin{center}
    \includegraphics[width=0.60\textwidth]{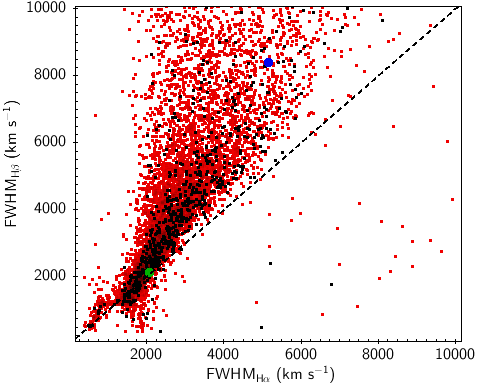}
  \end{center}
  \caption{FWHM of the broad components from the spectra of type 1 AGNs with detected H$\alpha$\ and H$\beta$ emission lines. Objects with S/N < 20 are shown as red points, while objects with S/N > 20 are marked in black. The black dashed line indicates the 1:1 relationship, the filled blue circle marks the position of 1RXS~J075111.5+174350, while the green circle represents Mrk~1243.}
\end{figure}
\section{Discussion}
The main motivation of our study is to extract emission-line parameters from spectroscopic databases through automated procedures designed to model different types of spectra. 
Since one of the fundamental parameters used to distinguish NLS1s from other objects involves the width of the H$\beta$\ broad-line component, we extracted the profiles of H$\beta$\ and H$\alpha$\ from a sample of 13283 SDSS spectra, where both emission lines were succesfully modelled. By plotting the relationship between the FWHM of H$\beta$\ and H$\alpha$, as shown in Fig.~2, we find that the two lines typically exhibit similar profiles in the low velocity regime, while H$\beta$\ appears to become increasingly broader than H$\alpha$\ in objects with larger velocity widths. This trend is clearly observable even when reducing the sample to the spectra with a S/N > 20 in the continuum at $5100\,$\AA. Differences in the profiles of these two lines, as well as in their reverberation lags, have already been reported and it has been suggested that they might be connected with changes in the line emissivity across the broad line region (BLR, \cite{Kaspi00, Bentz10}).
\begin{figure}[t]
  \begin{center}
    \includegraphics[width=\textwidth]{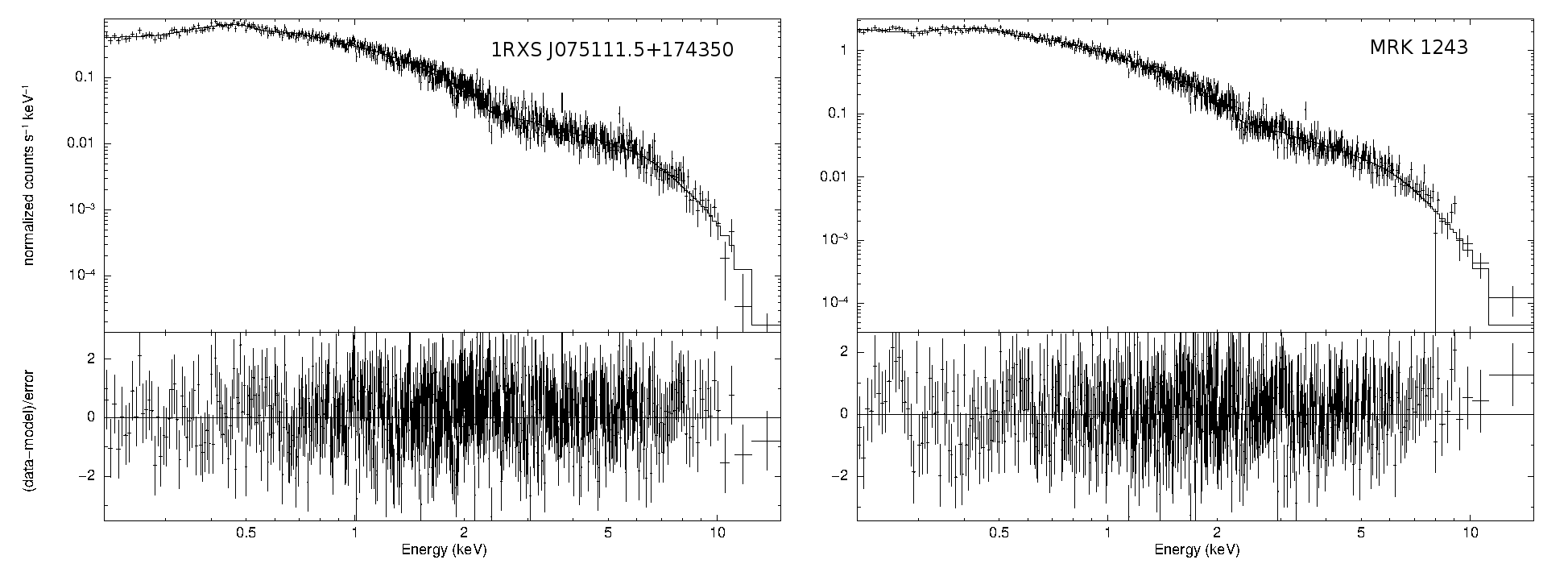}
  \end{center}
  \caption{{\bf Left panel - }{\it EPIC}-pn X-ray spectrum of 1RXS~J075111.5+174350, a broad lined AGN that exhibits a remarkable profile difference in the H$\alpha$\ and H$\beta$\ line profiles, with a corresponding X-ray absorption from an ionized medium exceding the foreground absorption of galactic neutral H. {\bf Right panel - }{\it EPIC}-pn X-ray spectrum of Mrk~1243, a narrow-lined Type 1 object with comparable line profiles and no evidence of absorption from intervening gas.}
\end{figure}
There is, however, also the possibility that a difference in the optical depth at the wavelength of the two lines may result in the observed effect. In particular, it is possible that a layer of ionized plasma, with recombining H atoms that cascade through the first excited level, may result in a different optical depth to the Balmer line photons, since the absorption coefficient for such atoms is $k_\lambda \propto \lambda^2$. In order to look for possible clues of such structures, we investigated the X-ray emission of the sources with fitted emission lines, matching our sample with XMM sources with an estimated flux larger than $10^{-12}\, {\rm erg\, cm^{-2}\, s^{-1}}$ in the 0.2 -- 12~keV range and having high S/N SDSS spectra. In Fig.~3, we show two examples of X-ray spectra:1RXS~J075111.5+174350, a broad-lined type 1 with remarkable line profile differences, and Mrk~1243, a narrow-lined Seyfert 1 with very similar line profiles. While in the first case, the X-ray spectrum shows the effect of absorption in excess of the Milky Way neutral H, with an estimated column density of $1.19 \times 10^{20}\, {\rm cm}^{-2}$, no excess absorption is detected in Mrk~1243.

The intervening medium observed in some AGNs may be the result of material ejected from the nuclear regions and -- lying along the line of sight to the central source, under the influence of its strong ionizing continuum -- it should be probably free of dust. The typically more similar profiles of narrow-line emitting sources, particularly in NLS1s, suggest that this type of effect is not similarly common in these objects. This property is consistent with other evidence which, based on IR observations, suggests that the environment of NLS1s is typically richer in dust than that of other type 1 objects \cite{Chen17}.

\section{Conclusions}
From analysis of the broad components of the emission lines in type 1 AGNs, we find that a systematic difference between H$\alpha$\ and H$\beta$\ becomes more and more significant with increasing line width. If this effect is related to the presence of a layer of ionized gas in front of the source, as some X-ray spectra appear to support, then narrow lined type 1 AGNs, and in particular NLS1s, might be operating in a phase where the development of this structure has not yet been fully accomplished. This consideration, which favours the evolutionary interpretation of the nature of NLS1s over a mere geometrical effect, is also consistent with the claim that NLS1s are hosted in dust-rich environments that would otherwise be swept by the nuclear activity. Clearly the evidence for such an interpretation is still limited and much more effort must be devoted to the execution of multiple wavelength investigations that may include sufficient optical, IR and X-ray data to better clarify the matter.

\section*{Acknowledgements}

This conference has been organized with the support of the
Department of Physics and Astronomy ``Galileo Galilei'', the 
University of Padova, the National Institute of Astrophysics 
INAF, the Padova Planetarium, and the RadioNet consortium. 
RadioNet has received funding from the European Union's
Horizon 2020 research and innovation programme under 
grant agreement No~730562. J. H. Fan's work is partially
supported by the National Natural Science Foundation of
China (NSFC 11733001, NSFC U1531245). We thank the referee
for comments and suggestions leading to the improvement of
this text.

%
\newcommand{\aap}{A\&A}
\newcommand{\aj}{AJ}
\newcommand{\apj}{ApJ}
\newcommand{\apjl}{ApJL}
\newcommand{\apjs}{ApJS}
\newcommand{\apss}{Ap\&SS}
\newcommand{\araa}{ARA\&A}
\newcommand{\memsai}{MemSAIt}
\newcommand{\mnras}{MNRAS}
\newcommand{\na}{NewA}
\newcommand{\pasa}{PASA}
\newcommand{\pasp}{PASP}
\bibliographystyle{JHEP}
\bibliography{procLaMura}

\end{document}